\documentstyle[twoside,fleqn,styleforqcd]{article}
\def\greaterthansquiggle{\raise.3ex\hbox{$>$\kern-.75em\lower1ex\hbox{$\sim$}}}
\def\lessthansquiggle{\raise.3ex\hbox{$<$\kern-.75em\lower1ex\hbox{$\sim$}}}
\newcommand{\bdi}{\begin{displaymath}}
\newcommand{\edi}{\end{displaymath}}
\newcommand{\bfi}{\begin{figure}}
\newcommand{\efi}{\end{figure}}

\newcommand{\beq}{\begin{equation}}
\newcommand{\eeq}{\end{equation}}

\newcommand{\gaf}{\gamma_{5}}

\newcommand{\beqa}{\begin{eqnarray}}
\newcommand{\eeqa}{\end{eqnarray}}

\newcommand{\ra}{\rightarrow}

\newcommand{\wt}{\widetilde}

\newcommand{\dsla}{\partial\hspace{-6pt} /  }  
\newcommand{\Asla}{A\hspace{-6.5pt}  /  }

\title{Decay widths and scattering processes in massive QED$_2$
\thanks{Work supported by a Schr\"odinger Stipendium of the Austrian FWF}}
\author{C. Adam \\ {~~~} \\
         Center for Theoretical Physics,
         Massachusetts Institute of Technology \\
         Cambridge, Massachusetts 02139 \\
         {~~~~~~~~~~~~~~~~~~~~~~~~~~~~~~~~~~~~~~~~~~~} 
         (MIT-CTP-2661, {~~~~~} September 1997)}

\begin{document}

\begin{abstract}
Using mass perturbation theory, we infer the bound-state spectrum of
massive QED$_2$ and compute some decay widths of unstable bound
states. Further, we discuss scattering processes, where all the resonances and 
particle production thresholds are properly taken into account by our
methods.

\end{abstract}

\maketitle

\section{Introduction}

\input psbox.tex
\let\fillinggrid=\relax

QED$_2$ with one massive fermion,
\beq
L=\bar \Psi (i\dsla -e\Asla +m)\Psi -\frac{1}{4}F_{\mu\nu}F^{\mu\nu} ,
\eeq
is an interesting toy model for the study of QCD-like properties 
\cite{CJS}-\cite{MSMPT} (see \cite{MSMPT} for a more extensive list of 
references). It 
shares with QCD properties like the occurrence of instantons and a
nontrivial $\theta$ vacuum, the presence of a fermion condensate and
confinement of the fundamental fermion of the theory -- the spectrum
solely consists of mesons. More precisely, we will find two stable 
particles in the model, a lightest ``Schwinger boson'', corresponding to
the $\eta'$ of QCD, and a second particle that may be thought of as a
bound state of two Schwinger bosons. Further, there exist unstable
higher bound states that may decay into the two stable particles and
behave like resonances in scattering cross sections. 
Using mass
perturbation theory, we will perform
a resummation of the $n$-point functions that will prove essential for
our results. Further, we will describe the general bound state structure of
the model and show how to compute decay widths of the unstable bound
states and scattering cross sections of the stable particles. 

The massless model, which is the starting point for mass perturbation
theory, may be solved exactly \cite{Sc1,LS1}. 
It is equivalent to the theory of a free,
massive boson field $\Phi$, where $\Phi$ is related to the vector current,
$J_\mu =:(1/\pi)^{1/2} \epsilon_{\mu\nu}\partial^\nu \Phi$. The two-point
function of $\Phi$ is just the massive scalar boson propagator
$D_{\mu_0}(x-y)$, 
$\mu_0 =e/\sqrt{\pi}$.
Further VEVs that may be computed are VEVs of chiral densities
$S_\pm =\bar\Psi (1/2) ({\bf 1}\pm \gaf)\Psi$,
\bdi
\langle S_{H_1}(x_1)\cdots S_{H_n}(x_n)\rangle_0 = 
\edi
\beq
e^{ik\theta} 
\Bigl( \frac{\Sigma}{2}\Bigr)^n \exp
\Bigl[ \sum_{i<j}\sigma_i \sigma_j 4\pi D_{\mu_0} (x_i -x_j)\Bigr] 
\eeq
where $\sigma_i =\pm 1$ for $H_i =\pm$,
$\theta$ is the vacuum angle and $k=n_+ -n_-=\sum \sigma_i$ is the instanton
number of the contributing instanton sector. 
Further $\Sigma$ is the fermion condensate for $\theta =0$,
$\Sigma =\langle \bar\Psi \Psi \rangle_0^{\theta =0}$.

Now the mass perturbation theory may be traced back to space-time integrations 
of these chiral VEVs, e.g. ($S=S_+ +S_-$)
\beq
Z(m,\theta)= \langle 
\sum_{n=0}^\infty \frac{m^n}{n!}\prod_{i=1}^n \int dx_i \bar\Psi (x_i) \Psi
(x_i) \rangle_0 ,
\eeq
\beq
\langle \hat O \rangle_m =\frac{1}{Z(m,\theta)} \langle \hat O
\sum_{n=0}^\infty \frac{m^n}{n!}\prod_{i=1}^n \int dx_i S
(x_i) \rangle_0 .
\eeq
However, as $\exp (\pm 4\pi D_{\mu_0}(x)) 
\stackrel{x\to\infty}{\longrightarrow} 1$, 
the perturbation expansion, as it stands, is IR divergent.
Therefore, one has to expand the exponentials  into the functions 
$
E_\pm (x)=e^{\pm 4\pi D_{\mu_0}(x)}-1
$.
When all expressions are reexpressed in $E_\pm$, it may be shown that the
vacuum energy is extensive, $Z(m,\theta)=\exp (V\epsilon(m,\theta))$,
($V$\ldots space-time volume, $\epsilon(m,\theta)$\ldots vacuum energy
density), and, therefore, all VEVs are finite for $V\to\infty$.

Further, because $S=S_+ +S_-$ occurs in the perturbation expansion (3,4) and
$S_+$ and $S_-$ have slightly different VEVs (2), the Feynman rules
of the mass perturbation theory acquire a matrix structure.
E.g. the propagator connecting two ``interaction vertices'' $mS$ is given 
by the matrix
\beq
{\cal E}(p) =\left( \begin{array}{cc}\wt E_+ (p) & \wt E_- (p) \\ \wt E_- (p)
 & \wt E_+ (p) \end{array} \right)
\eeq
where the individual $++$, $+-$, etc. entries connect the $S_+ S_+$, $S_+ S_-$,
etc. chiral densities, see (2,3). Because of (2) an arbitrary number of 
propagators (5) may meet at one vertex, and each vertex is an $n$-th rank
tensor ${\cal G}$ when $n$ propagators meet there \cite{MSMPT}. 
Only two components of this tensor
are nonzero,
\beq
{\cal G}_{++\cdots +} =g \; ,\quad 
{\cal G}_{--\cdots -} =g^* \; ,\quad g=m\frac{\Sigma}{2}e^{i\theta}
\eeq
(corresponding to $S=S_+ +S_-$). The graphical Feynman rules are given
in Fig. 1.

$$\psannotate{\psboxscaled{400}{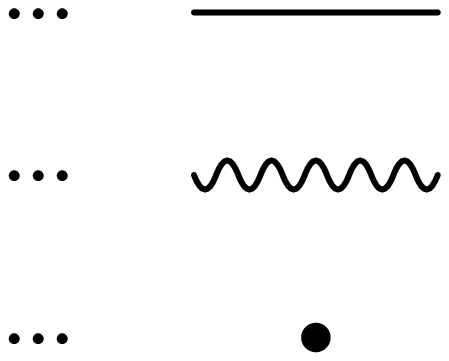}}{\fillinggrid
\at(0.7\pscm;-0.5\pscm){Fig. 1} \at(-1.9\pscm;1.1\pscm){${\cal G}$}
\at(-2.6\pscm;2.8\pscm){${\cal E}(p)$} \at(-3\pscm;4.5\pscm){$\wt
D_{\mu_0}(p)$}}$$

\vspace{0.3cm}

\section{$n$-point functions, bound-state structure, decay widths}

Using the Feynman graphs of Fig. 1, we find for the bosonic two-point function,
after amputation of the external boson lines, Fig. 2, where we introduced
the exact propagator $\Pi$ that is defined in Fig. 3.

$$\psannotate{\psboxscaled{400}{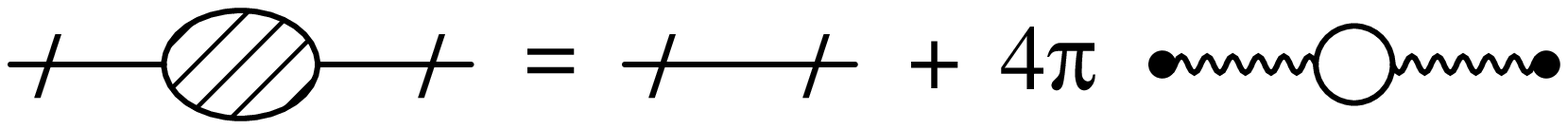}}{\fillinggrid
\at(6.9\pscm;-0.5\pscm){Fig. 2}}$$

$$\psannotate{\psboxscaled{350}{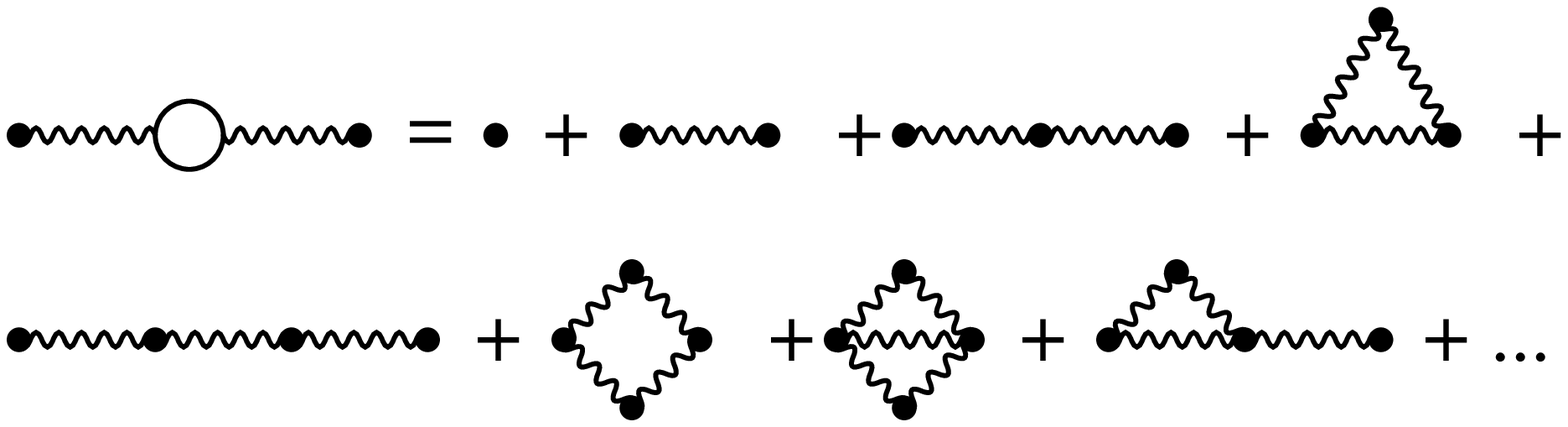}}{\fillinggrid
\at(8.9\pscm;-0.5\pscm){Fig. 3}}$$

\vspace{0.1cm}
 
The expression corresponding to Fig. 3 reads 
\beq
{\cal G}\Pi (p) := {\cal G} +{\cal G}{\cal E}(p){\cal G}+{\cal G}{\cal E}(p)
{\cal G}{\cal E}(p){\cal G} +\ldots 
\eeq
In (7) there occur two types of terms: terms that factorize in momentum space
(their graphs fall into two pieces when they are cut at a vertex); the other
type terms we call non-factorizable (n.f.) (they are analogous to 1PI graphs
of other theories). Each term in $\Pi$ is a n.f. part times an arbitrary
part of $\Pi$, therefore it holds that
$
\Pi (p)={\bf 1} +\Pi^{\rm n.f.}(p)\Pi (p)
$,
which may be solved,
\bdi
\Pi (p)=
\edi
\beq
\frac{1}{\det ({\bf 1}-\Pi^{\rm n.f.}(p))} 
\left( \begin{array}{cc} 1-\Pi^{\rm n.f.}_{--} & 
\Pi^{\rm n.f.}_{+-} \\ \Pi^{\rm n.f.}_{-+} &
1-\Pi^{\rm n.f.}_{++} \end{array} \right)
\eeq
where, in leading order, $\Pi^{\rm n.f.}(p)$ is just ${\cal E}(p){\cal G}$,
$\Pi^{\rm n.f.}_{++}(p)=g\wt E_+ (p)+o(g^2)$, etc. We will be especially 
interested in the determinant in the denominator of (8), because the zeros 
of its real part will give us the mass poles of the theory, whereas the
imaginary parts at the mass poles lead to the decay widths of the unstable
bound states.

In an analogous fashion higher $n$-point functions may be reexpressed in
terms of exact propagators and n.f. higher $n$-point functions. We show
the graphs for the 3- and 4-point functions in Figs. 4, 5 (the triangle
and quadrangle denote the n.f. 3- and 4-point functions, respectively).

$$\psannotate{\psboxscaled{250}{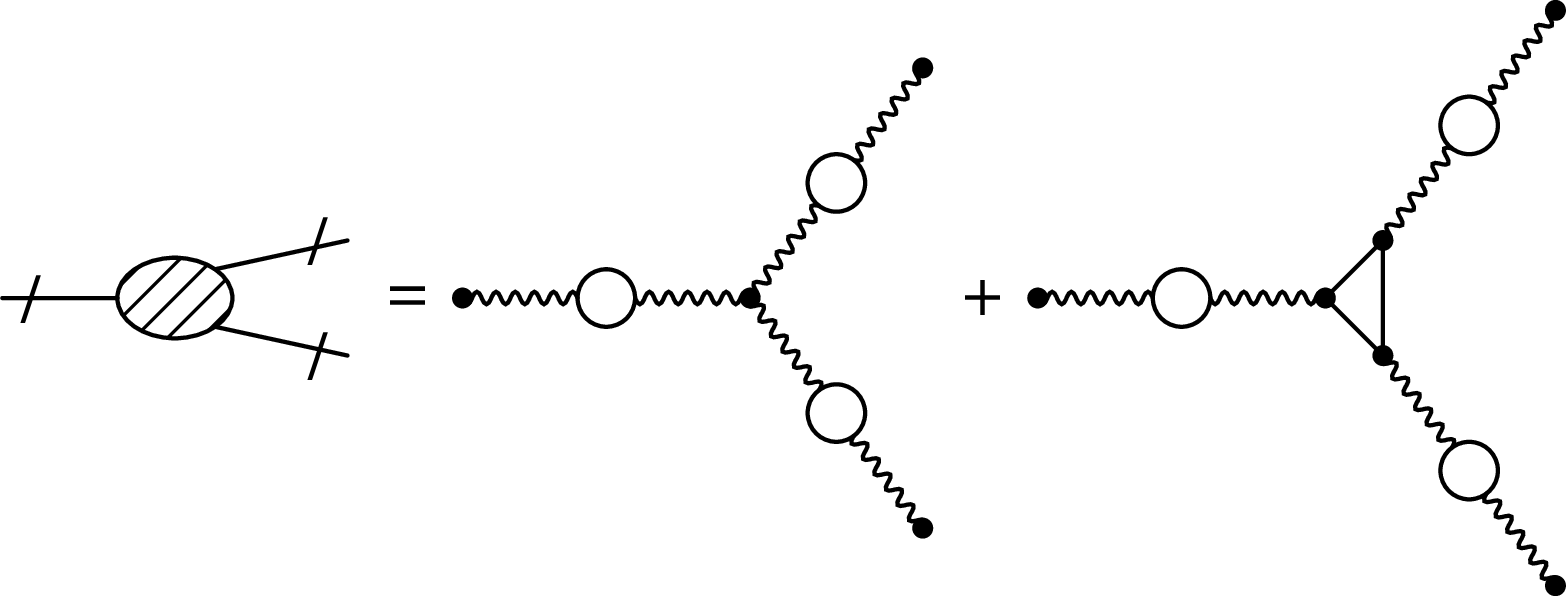}}{\fillinggrid
\at(10.9\pscm;-0.5\pscm){Fig. 4}}$$

$$\psannotate{\psboxscaled{250}{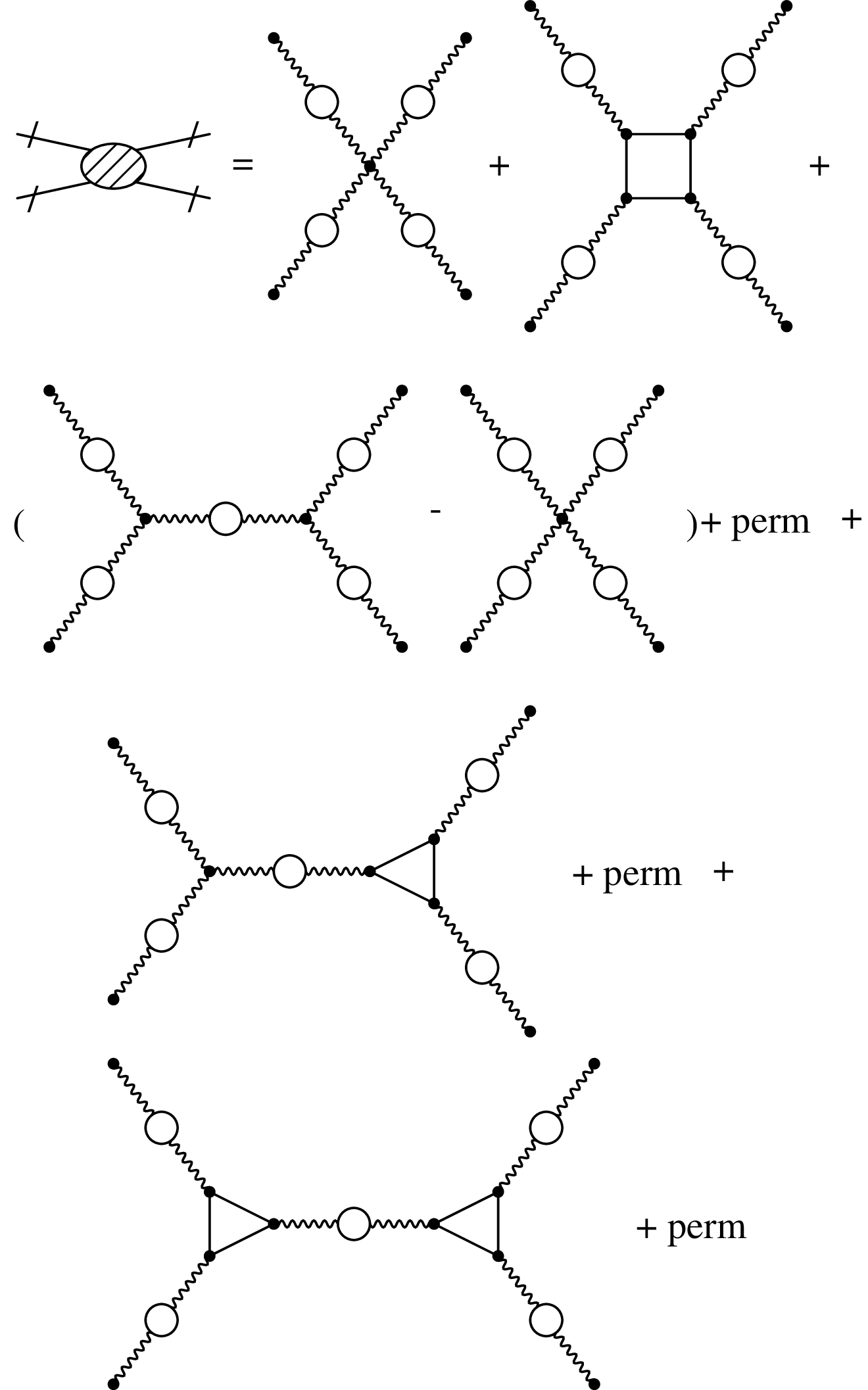}}{\fillinggrid
\at(10.9\pscm;-0.5\pscm){Fig. 5}}$$ 

As stated, the determinant in (8) will lead to the bound states and decay 
widths, therefore let us investigate it more closely.
\bdi
N(p):=\det ({\bf 1}-\Pi^{\rm n.f.}(p))=
\edi
\beq
1-m\Sigma\cos\theta \wt E_+ (p)
+o(m^2)
\eeq
\beq
\wt E_\pm (p)=:\sum_{n=1}^\infty d_n (p)\; ,\; d_n (p):=\frac{(4\pi)^n}{n!}
\wt{D^n_\mu}(p)
\eeq
The $d_n(p)$ are just $n$-boson blobs (up to a factor) and behave like follows.
They are singular at $p^2 =(n\mu)^2$ (real particle production threshold), 
therefore they are large enough slightly below to balance the small $m
\Sigma\cos\theta$ in (9) and make the whole expression (9) vanish. Further
they acquire an imaginary part above the threshold. As a consequence we
find mass poles slightly below all $n$-boson thresholds in (8) ($n$-boson
bound states). At the lowest bound-state mass $M_2^2=(2\mu)^2 -\Delta_2$,
$N(p)$ has no imaginary part, therefore the lowest bound state $M_2$ is
stable, like the fundamental boson $\mu$. For higher mass poles
$M_n^2=(n\mu)^2 -\Delta_n$ the $d_i$ have imaginary parts for $i=2,\ldots
,n-1$, therefore the $n$-boson bound state may decay into $2, \ldots,n-1$
fundamental bosons.

However, there is still something missing. As just stated, the $M_2$ is a 
stable particle, therefore it should be possible in a final state, so
where is it? For an answer we need a further resummation.
The $M_2$ mass pole and propagator is found, in lowest order, by the
graphs of Fig. 6. So let us include into $N(p)$ all the graphs of Fig. 7.

$$\psannotate{\psboxscaled{400}{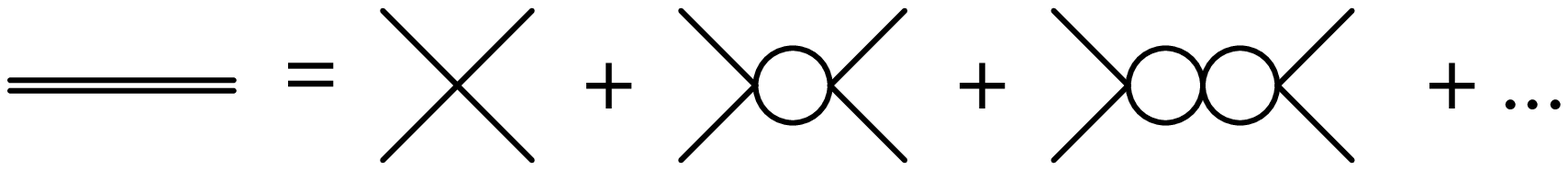}}{\fillinggrid
\at(8\pscm;-0.5\pscm){Fig. 6}}$$

$$\psannotate{\psboxscaled{400}{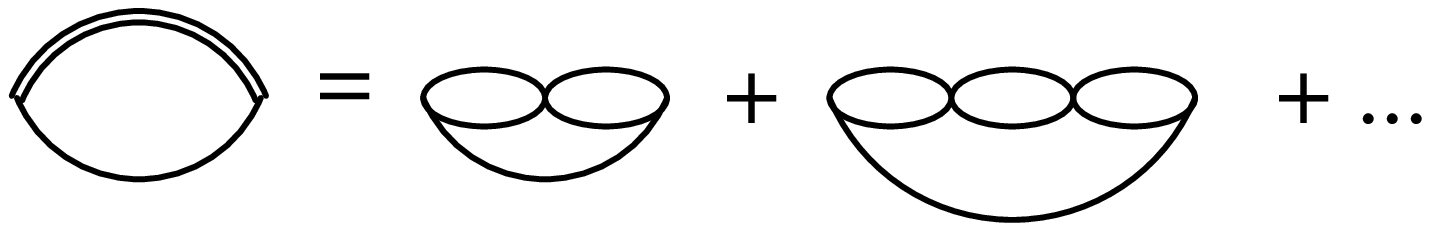}}{\fillinggrid
\at(5.9\pscm;-0.5\pscm){Fig. 7}}$$

Each graph in Fig. 7 is n.f., therefore they all contribute to $N(p)$.
Further, Fig. 6 is a perfectly legitimate propagator of a stable particle, 
therefore Fig. 7 is a perfectly legitimate two-boson blob consisting of
one $\mu$ and one $M_2$. We call this blob $d_{1,1}(p)$ and may find by
analogous reasoning e.g. a blob of two $M_2$, $d_{2,0}$, etc. 
Therefore we write for the further resummed $N(p)$
\beq
N(p)=1-m\Sigma\cos\theta\sum_{m,n}d_{m,n}(p)
\eeq
and the additional $d_{m,n}$ have properties completely analogous to those
discussed for the $d_n$ above. As a consequence we find bound states
$M_{m,n}^2 =(mM_2 +n\mu)^2 -\Delta_{m,n}$ slightly below all thresholds
and possible decays into all combinations of $\mu$ and $M_2$ that are
allowed kinematically.

Now the bound-state masses may be computed explicitly by solving the equation
${\rm Re}\, N(p)=0$ near the individual thresholds 
(see e.g. \cite{GBOUND,MSMPT}). 
In the vicinity of the mass poles ${\rm Re}\, N$ may be approximated
by the first Taylor coefficient,
\bdi
N(p^2 \sim M^2_{m,n})\sim 
\edi
\beq
c_{m,n}(p^2 -M^2_{m,n})+i\, {\rm Im}\, N(M^2_{m,n})
\eeq
$c_{m,n}$ may be computed from the respective binding energy of the
bound state $M_{m,n}$ (see \cite{MSMPT}), and is, of course, related to
the residue of the exact propagator $\Pi$ at the mass pole $M^2_{m,n}$. $
{\rm Im}\, N(M^2_{m,n})$ consists of all ${\rm Im}\, d_n$ with thresholds below
$M^2_{m,n}$, and these ${\rm Im}\, d_n$ are wellknown kinematical functions.
E.g. for the $M_3$ three-boson bound state we find ($s:=p^2$)
\bdi
N(s\sim M_3^2)=c_3 (s-M_3^2)-
\edi
\beq
im\Sigma\cos\theta 
(\frac{\sin^2\theta}{\cos^2\theta}{\rm Im}\, d_2 (M^2_3) +{\rm Im}\, 
d_{1,1}(M_3^2))
\eeq
corresponding to the two partial decay channels $M_3\to 2\mu$ and $M_3\to
M_2 +\mu$. The ${\rm Im}\, d_2$ term has an additional $\theta$ factor, because
the decay $M_3\to 2\mu$ is parity forbidden.
Comparing with the general expression for the total decay width $\Gamma$ 
of a bound state $M$,
\beq
\frac{1}{N(s\sim M^2)}\sim \frac{\rm const.}{s-M^2 -i\Gamma M} 
\eeq
we find after inserting all the numbers
\beq
\Gamma_{M_3\to 2\mu}\simeq  3.6 \mu\frac{\sin^2\theta}{\cos^2\theta}\exp 
(-0.93 \frac{\mu}{m\cos\theta}) 
\eeq
\beq 
\Gamma_{M_3\to M_2 +\mu}\simeq  44\mu \exp (-0.93\frac{\mu}{m\cos\theta})
\eeq
Analogously we may find for the lightest unstable bound state $M_{1,1}$
\beq
\Gamma_{M_{1,1}\to 2\mu}\simeq 21340 \mu \Bigl( \frac{m\cos\theta}{\mu}\Bigr)^5
\frac{\sin^2\theta}{\cos^2\theta}
\eeq
Here only one decay channel exists, and the decay is parity forbidden
(see \cite{DECAY,MSMPT} for details).

\section{Scattering processes}

A general elastic scattering cross section is
\beq
\sigma_{ab\ra ab}=\frac{C_{\rm sym}|{\cal M}(s)|^2}{2w^2 (s,M_a^2,M_b^2)}
\eeq
where the final state symmetry factor $C_{\rm sym}=1/(n!)$ for each $n$
identical particles in the final state, ${\cal M}$ is the transition
matrix element and $w^2 (x,y,z)=(x^2 +y^2 +z^2 -2xy-2xz-2yz)$ is a wellknown
kinematical function.

For lowest order two-particle elastic scattering we need the first graph
on the r.h.s. of Fig. 5. There 4 exact propagators meet at one vertex,
and each exact propagator may describe a $\mu$ or $M_2$, because these
are the two stable particle poles of $\Pi$. Choosing e.g. $2\mu \ra
2\mu$ elastic scattering for definiteness we obtain in lowest order
\beq
\sigma_{2\mu \ra 2\mu}(s)=\frac{r_1^4 \frac{1}{2}(m\Sigma\cos\theta)^2}{2
w^2 (s,\mu^2 ,\mu^2)}
\eeq
where $r_1 =4\pi ={\rm Res}\, \Pi (\mu^2)$ is the residue of the exact 
propagator at the first mass pole $\mu$. 

The first and resummed second order $s$-channel contribution is given by 
Fig. 8, see Fig. 5.

$$\psannotate{\psboxscaled{300}{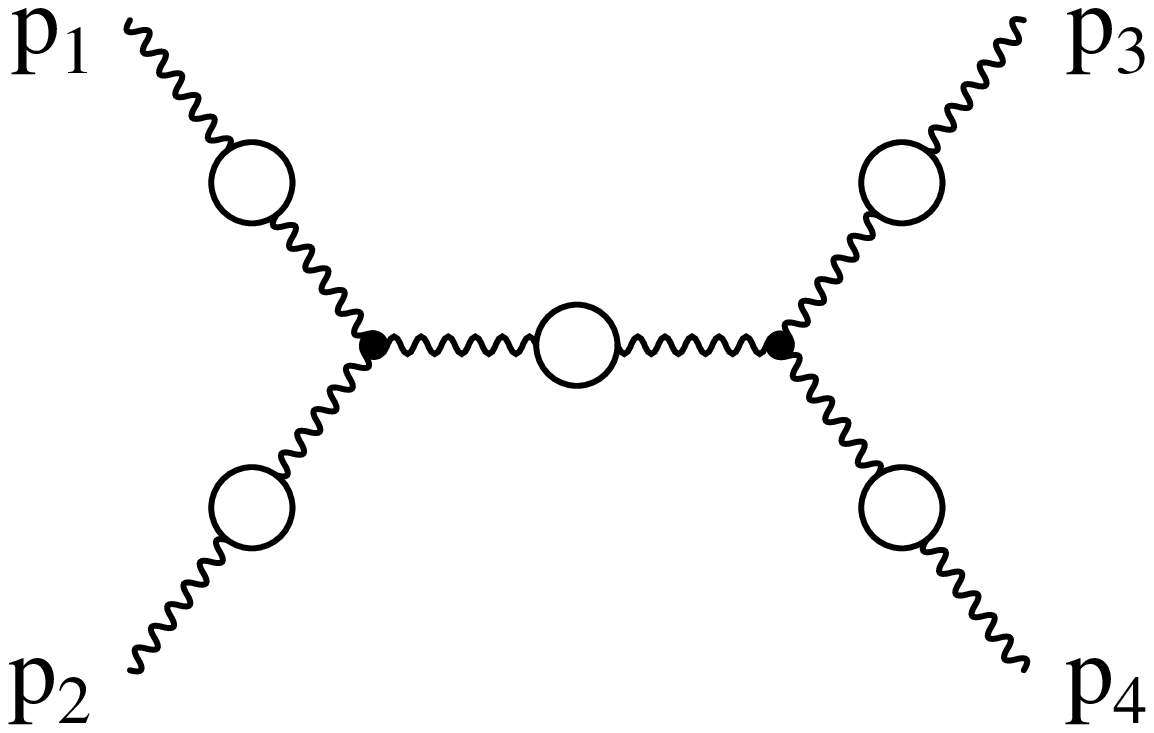}}{\fillinggrid
\at(5.7\pscm;0\pscm){Fig. 8}}$$

We again fix the initial state to be $2\mu$, and we allow
for an arbitrary final state, because we want to use the optical theorem
in the sequel,
\beq
\sigma^{\rm tot}_{2\mu\to f}(s)=\frac{r^2_1}{w(s,\mu^2 ,\mu^2)}
{\rm Im}\, {\cal M}_{2\mu \to 2\mu} (s)
\eeq
Observe that only initial state factors occur explicitly in (20), therefore $
{\rm Im}\, {\cal M}$ must produce all the final state factors. Further
we will restrict to the case $\theta=0$ in the sequel, because it is easier.
In this case ${\cal M}$ reads
\bdi
{\cal M}^{\theta =0}_{2\mu \to 2\mu}=
\frac{m\Sigma}{1-m\Sigma (\wt E_+ -\wt E_-)}
\edi
\beq
=\frac{m\Sigma}{1-m\Sigma (d_2 +d_{2,0}+d_4 +\ldots )}
\eeq
i.e. only parity even blobs occur. Inserting this into (20) we find
\bdi
\sigma^{{\rm tot},\theta=0}_{2\mu \to f}(s)=\frac{r_1^2 m^2 \Sigma^2}{
w(s,\mu^2,\mu^2)}\cdot
\edi
\beq
\cdot \frac{{\rm Im}\, d_2 +{\rm Im}\, d_{2,0} +{\rm Im}\, d_4 +\ldots}{
[1-m\Sigma ({\rm Re}\, d_2 +\ldots)]^2 +m^2\Sigma^2 ({\rm Im}\, d_2
+\ldots )^2}
\eeq
We indeed recover the required final state factors. Remember that
${\rm Im}\, d_n =(r_1^n /n!) {\rm Im}\, \wt{D^n_\mu}$. Here ${\rm Im}\,
\wt{D^n_\mu}$ is the final state with the phase space integrations, $r_1^n$
are the residues of the final state $\mu$, 
and $1/n!$ is the final state symmetry factor. 
Further, it is obvious from Fig. 8 that $M_2$ may be present in the final 
state. Therefore, it must be present in the intermediate state, too, in
order to saturate the optical theorem, i.e. our inclusion of the
general $d_{m,n}$ into $N(p)$ in the last section
is absolutely crucial for unitarity.

Finally we want to evaluate $\sigma^{{\rm tot},\theta =0}_{2\mu \to f}(s)$
for some specific values of $s$. For $(2\mu)^2 <s<M_{2,0}^2$
the $m\Sigma$ terms in the denominator
of (22) are small compared to 1, and further only ${\rm Im}\, d_2 (s)\ne 0$.
As a consequence, (22) reduces to the first order result (19). 
Next we 
investigate (22) at the first bound-state mass, $s=M_{2,0}^2$. There the
term $1-m\Sigma {\rm Re}\, (\ldots)$ in the denominator vanishes by
definition, and we find a $m^2\Sigma^2 ({\rm Im}\, d_2)^2$ in the denominator,
a $m^2\Sigma^2 {\rm Im}\, d_2$ in the numerator, and a further term 
proportional to ${\rm Im}\, d_2$ in the numerator from the initial state
factor $w(M_{2,0}^2,\mu^2,\mu^2)$. I.e. everything cancels and we are left
with $\sigma^{{\rm tot},\theta =0}_{2\mu \to f}(M^2_{2,0})=4$. This is much
larger than the first order result, i.e. a resonance occurs.
At the first higher production threshold $s=(2M_2)^2$ the term ${\rm Im}\,
d_{2,0}$ is singular. It occurs linearly in the numerator and quadratically
in the denominator, therefore the scattering cross section (22) vanishes.
In addition, the $2\mu \to 2M_2$ inelastic scattering channel opens at
$s=(2M_2)^2$.

For higher $s$ the above pattern repeats. Therefore we find the following
general behaviour. Far away from all thresholds and bound states the
scattering cross section is well described by the lowest order result (19).
At the position of a bound state a resonance occurs, and the resonance widths
are related to the binding energies, because the cross section goes down
at the corresponding particle production threshold (a more detailed discussion
may be found in \cite{SCAT,MSMPT}).

\section{Summary}

We have uncovered quite a rich physical structure in the course of our 
investigation. We found two
stable particles in the theory, namely the Schwinger boson $\mu$ and the
two-boson bound state $M_2$. Higher (unstable) bound states may be formed 
out of an arbitrary number of $\mu$ and $M_2$. Further, these unstable 
bound states may decay into all combinations of $\mu$ and $M_2$ that are
possible kinematically. 

 For scattering processes we found that far from all resonances and particle 
production thresholds the scattering cross section is well described by
a lowest order computation. 
Whenever it is near a bound-state mass, 
the scattering cross section has a local maximum, i.e. 
a resonance occurs. Moreover, for all energies where a new final state 
becomes possible kinematically, the corresponding real particle production 
threshold indeed occurs.

\end{document}